\documentclass[twocolumn]{webofc}
\usepackage[varg]{txfonts}   % Web of Conferences font

\usepackage{hyperref}
\begin{document}
\title{The transition from Galactic to extragalactic cosmic rays: the high--energy end of the Galactic spectrum}

\author{\firstname{Pierre} \lastname{Cristofari}\inst{1}\fnsep\thanks{\email{pierre.cristofari@obspm.fr}}
}

\institute{Laboratoire Univers et Th\'eories, Observatoire de Paris, Universit\'e PSL, Université Paris Cité, CNRS, F-92190 Meudon, France
     }

\abstract{%
Understanding the transition from Galactic to extragalactic cosmic rays (CRs) is essential to make sense of the Local cosmic ray spectrum. Several models have been proposed to account for this transition in the 0.1 - 10 $\times 10^{18}$ eV range. For instance: ankle models, where the change from a steep Galactic component to a hard extragalactic spectrum occurs in the $4-10 \times 10^{18}$ eV region, dip models, where the interactions of CR protons with the CMB producing electron-positron pairs shape the ankle, or mixed composition models, in which extragalactic CRs are composed of nuclei of various types.

In all these scenarios, the low-energy part of the transition involves the high-energy part of the Galactic component. Therefore, any information on the Galactic component, such as maximum energy, chemical composition, and spectrum after propagation, is crucial to understanding the Galactic-extragalactic transition.
We briefly review the high-energy part of the CR spectrum expected from the best potential sources of Galactic CRs.
}
\maketitle
\section{Introduction}
\label{intro}
\begin{figure}
\centering
\sidecaption
\includegraphics[width=8cm,clip]{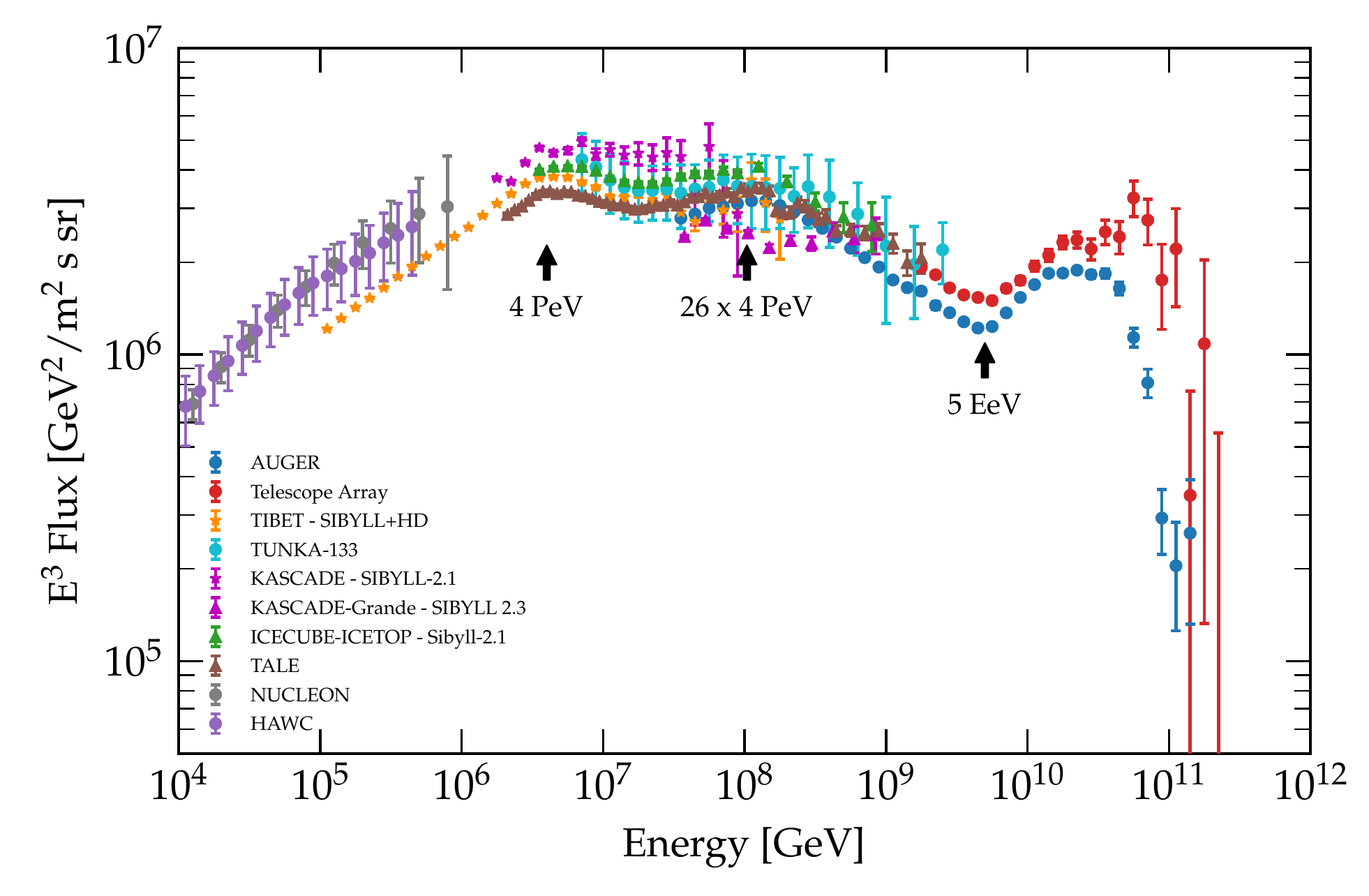}
\caption{The local interstellar proton spectra. Data compiled by C. Evoli \url{https://github.com/carmeloevoli}}
\label{fig:LIS}       % Give a unique label
\end{figure}

The local interstellar spectrum (LIS) of cosmic rays (CRs), compiling the measurements of dozens of experiments, represents the local flux of CRs at the of Earth's atmosphere as a function of the particle's kinetic energy. The most remarkable features in the LIS include at least the \textit{knee} and the \textit{second knee}, associated to a spectral steepening, and the \textit{ankle}, associated to a spectral hardening. 
For protons, accounting for about $\sim 90$ \% of CRs,  these features are located at energy  of $\sim 3-4$ PeV(=$10^{15}$ eV) (\textit{knee}), $\sim 1-5 \times 10^{17}$ PeV (\textit{second knee}) , and $\sim 5$ EeV(=$10^{18}$ eV) (\textit{ankle}) (Fig.~\ref{fig:LIS}). It is often claimed that at least below the \textit{knee}, CRs must be of Galactic origin, and above the \textit{ankle}, the origin must be extragalactic.

A simple argument for the Galactic origin of the bulk of CRs comes from observing gamma rays from the Galactic disk. If CRs fill the Galaxy, they are expected to interact with the ISM material, and an enhanced gamma-ray signal is expected from Galactic regions where the density is higher, such as the Galactic disk~\citep{hayakawa1952}. Such a gamma-ray emission from the Galactic disk has been detected~\citep{Fermi1}. Moreover, when the  Larmor radius of CRs becomes larger than the typical size of the Galactic halo $h$, CRs cannot be confined within the Galaxy, i.e., when $E> 10^{18} \left( \frac{h}{\text{kpc}} \right) \left( \frac{B}{\mu \text{G}}\right)$ eV. 

Between the \textit{knee} and the \textit{ankle}, the transition from Galactic to extragalactic CRs has been the subject of numerous discussions, where various scenarios have been explored over the years, e.g., the 1) dip models, where the ankle is the result of pair-production losses of extragalactic CR protons on the CMB photons~\citep{berezinksy2006}; 2) mixed composition models, where various nuclei shape the spectrum, and the ankle appears to be a transition energy, and a light extragalactic component below the ankle is usually required~\citep{allard2005,hooper2007,aloisio2011}. 3) ankle models, where the transition at the ankle originates from the intersection of a flat extragalactic component and  a very steep Galactic spectrum. The extragalactic component is usually assumed to be pure protons~\cite {waxman2000,demarco2005,globus2015,unger2015}. These various models are discussed in several excellent reviews~\citep{aloisio2012,parizot2014,kachelriess2019}, and here we focus on several open questions on the Galactic spectrum around the protons knee. 

\section{Galactic cosmic rays around the knee}
The minimum requirement for the sources of Galactic CRs (and thus of Galactic protons) is that they must: 1) sustain the measured CR power; 2) inject a CR spectrum compatible with the measured LIS; 3) reach the \textit{knee}, i.e., at least the PeV range, and thus be \textit{pevatrons} for a least some time of their evolution; hence the hunt for pevatrons~\citep{cristofari_review2021}. 

\subsection{The knee}
The actual value for the proton knee is still open to discussion, for example, with ARGO-YBL finding a value around $\sim 700$ TeV~\citep{bartoli2015} and KASCADE a value of $\sim 3-4$ PeV~\citep{antoni2005}.
Moreover, measurements have shown that the value of the knee increases with Z and that therefore, the knees for different nuclei should shape the LIS above the knee for protons. 

A Z-dependent knee is in agreement with what is expected from diffusive shock acceleration (DSA), the prominent first-order Fermi mechanism responsible for particle acceleration at strong shocks of SNRs, for which the maximum energy of the accelerated particle is expected to be rigidity dependent. 
In addition, it has been argued that very sharp cut-offs for the spectra of the different nuclei fail to produce a smooth LIS above the proton knee and that steep spectra above the various knees naturally lead to a smooth LIS above the proton knee~\citep{parizot2014}. 
%gamma rays 

\subsection{Gamma rays}
The search, identification, and characterization of pevatrons are one of the key science projects of most of the major gamma-ray facilities, such as  CTA~\citep{CTA}, LHAASO~\citep{LHAASO}, or SWGO~\citep{SWGO}. 
In the very-high-energy (VHE) domain (i.e., TeV=10$^{12}$ eV domain), two main mechanisms can efficiently produce gamma rays at Galactic sources: 1) a leptonic mechanism, the inverse Compton diffusion of accelerated electrons; and 2) a hadronic mechanism, the pion production/decay from accelerated nuclei interacting with the ISM. (The bremsstrahlung of accelerated electrons can also potentially produce gamma rays but is often subdominant). 

The shape of the hadronic spectrum is roughly mimicking the spectrum of parent protons, and can roughly be approximated by a power-law in energy, exponentially suppressed at a maximum energy $E_{\gamma}\sim E_{\rm p}/10$ where $E_{\rm p}$ is the maximum energy of the parent protons so that the detection of $\sim 100$ TeV gamma rays can help probe $\sim$ PeV protons.  The shape of the leptonic gamma-ray spectrum is expected to be slightly more curved, with a maximum energy limited by the Klein-Nishina suppression in the cross-section of interactions, drastically impinging the efficient production of gamma--rays above $\gtrsim 50$ TeV. 

Most of the current VHE observatories, especially Imaging Atmospheric Cherenkov Telescopes (IACTs), such as H.E.S.S., MAGIC, and VERITAS are typically optimized in the $\sim$ 1-10 TeV range~\citep{chadwick2021}, a range in which differentiating between the two mechanisms is highly uneasy. For long, it was, however claimed that instruments optimized in the 100 TeV range could help directly differentiate between a hadronic and leptonic origin for the gamma rays since only the hadronic mechanism could efficiently produce gamma rays. And, at the same time, since $E_{\gamma}\sim E_{\rm p}/10$, the detection of 100 TeV would testify of the presence of $\sim$ PeV protons. It is now clear that things are slightly more complicated and that this former claim must be revised. For instance, the detection of the first dozen of Galactic pevatrons by LHAASO~\citep{cao2021} showed that for most of these sources, it is not clear which astrophysical object, what acceleration mechanism(s) and what emission mechanism(s) drive the gamma-ray emission. Several of these pevatrons are indeed expected to be leptonic pevatrons~\citep{deona2022} -- and at least the Crab is a commonly accepted example of a leptonic pevatron. The current difficulty of interpreting the origin of the gamma rays is especially due to the current performances of LHAASO: the exceptional sensitivity of the instrument in the 100 TeV range, which integrates the spectrum of all gamma-ray photons from the whole sky at all times,  is counterbalanced by a limited angular and spectral resolution, which will be eventually complemented for forthcoming instruments. 

Additionally, it has also been shown theoretically that hard electron spectra, in an environment allowing for fast inverse Compton cooling, can lead to hard gamma-ray spectra in the $\sim 100$ TeV range so that even the detection of a hard spectrum in the 100 TeV range could not firmly testify of a hadronic origin~\citep{vannoni2009,breuhaus2021}.

\subsection{The hunt for pevatrons}

Therefore, it seems that from the question "Which source(s) can be pevatrons?", we are now moving towards more specific questions: 
\begin{itemize}
\item Which source(s) can produce a hard-enough spectra above 10$^{15}$ eV? 
\item Which source(s) can produce a sufficient amount of PeV protons? 
\item Which Galactic source(s) can be \textit{super}pevatrons\footnote{I heard F. Aharonian use this term in a very nice talk in 2021, but not sure it is yet widely used}, i.e., accelerate particles up to rigidities $10^{16}$ V? 10$^{17}$ V?
\end{itemize}

\section{The case of supernova remnants}

Supernova remnants (SNRs) have for long been pointed as the most probable sources of Galactic CRs: because they can sustain the power of Galactic CRs; because they can roughly account for the injection spectrum $\propto E^{-2.1 ... -2.4}$ needed to explain the LIS, thanks to diffusive shock acceleration; and because they can theoretically be pevatrons. 
The maximum energy of particles accelerated at SNRs can be estimated with the Hillas criterion~\citep{hillas2005}: $E_{\max} \approx \xi \left( \frac{R_{\rm sh}}{\text{pc}}\right) \left( \frac{u_{\rm sh}}{\text{1000 km/s}}\right)  \left( \frac{B}{\mu \text{G}} \right) \text{TeV}$. With typical values of the shock speed and radius of an SNR, the magnetic field needs to reach values significantly larger the one of the ISM in order to reach the PeV range. Such values have been inferred from observations in the X-ray range, up to 2-3 orders of magnitude above the ISM~\citep{vink2012}. Theoretically, several mechanisms can lead to the amplification of the magnetic field at SNR shocks~\citep{gabici2016}. The mechanism expected to dominate for young SNRs is the \textit{non-resonant streaming of CRs}~\citep{bell2004} (also called the \textit{Bell} mechanism), due to the streaming of accelerated particles upstream of the shock exciting instabilities in the plasma on scales smaller than the Larmor radius of the streaming particles. The maximum energy of accelerated particles assuming magnetic field amplification alla Bell~\citep{schure2013,schure2014} is typically: $E_{\rm max} \approx \frac{R_{\rm sh}}{10} \frac{\xi c e \sqrt{4 \pi \rho (t) }}{\Lambda} \left(  \frac{u_{\rm sh}}{c} \right)^2$, where $\Lambda = \ln(p_{\rm max}/p_{\rm min})$, and can a priori for some remnants of unusual and rare core-collapse supernovae reach the PeV range. Moreover, it appears that if some SNRs can be pevatrons for some time of their evolution, then in order to not overproduce the total proton spectrum, the rate of these SNe has to be substantially lower than the typical Galactic rate, i.e. of the order $\sim 1-5$ \% $ \times 2-3/$century~\citep{cristofari2020}. 

It seems therefore possible that SNRs produce the bulk of CRs, with only a few of them contributing up to the knee. If only focussing on protons, several questions remain: 
\begin{enumerate}
\item The spectra of particles escaping from SNRs is expected to ber exponentially suppressed at high-energy: how can such suppression produce the transition from $\propto E^{-2.7}$ to $\propto E^{-3}$ at the knee? 
\item How do the different types of SNRs overlap to produce the very neat spectrum observed up to the knee? 
\item How can the deviation from the perfect power-law, such as the bump at $\sim 10$ TeV measured by DAMPE, be explained?  
\end{enumerate}

For question 1), as for any change of spectral index, the most probable hypotheses are: a break due to transport properties; two (or several) classes of sources overlapping; a steepening/break in the injection term from sources. The latter is probably the most reasonable hypothesis, as it has been shown that several effects, for instance, losses suffered by particles trapped inside an SNR, the leakage of particles in the ISM, could help diverge from a pure exponential suppression at high energy, and thus produce a steepening in the injected spectrum in the VHE range. For alternative candidates, clusters of massive stars, it has been shown that several effects, such as losses, escaping particles, or different diffusion regimes can substantially affect the slope at VHE~\citep{morlino2021}.

For question 2), the question remains very open: several quantities, such as the efficiency of particle acceleration,  the duration of the active phase of acceleration, the level of magnetic field downstream, can significantly affect the amount of particle and the spectra released by different SNRs in the ISM. It had already been pointed out that considering several populations of SNRs can naturally lead to bumps and features in the overall LIS proton spectrum~\citep{ptuskin2010}.
It was additionally shown that in order to reach the PeV range, very peculiar SNRs had to be considered, with, for instance, high mass-loss rates $\dot{M}\gtrsim 10^{-4}$ M$\odot$/yr, high total explosion energy $E_{\rm SN} \gtrsim 5 \times 10^{51}$ erg, and low-mass ejecta, typically $M_{\rm ej} \lesssim 2-3$ M$\odot$. For such objects, the total level of protons can naturally come close to the LIS. 

\begin{figure}
\centering
\sidecaption
\includegraphics[width=8cm,clip]{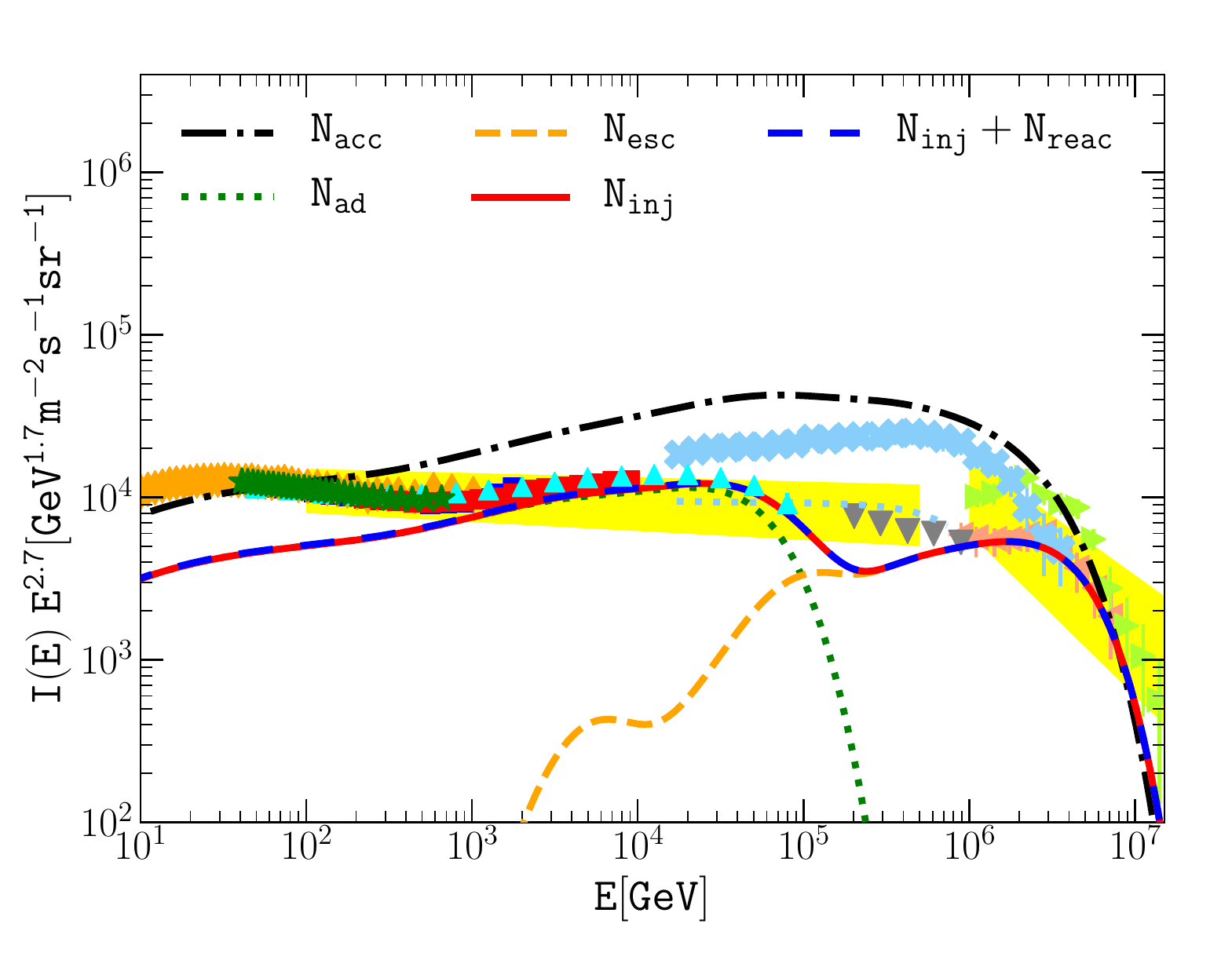}
\caption{Proton spectrum from a core-collapse SN with total explosion energy $E_{\rm SN}=6 \times 10^{51}$ erg, Mass-loss rate $\dot{M}= 10^{-5}$ M$_{\odot}$/yr and efficiency $\xi=5$ \%, the SN rate is $\nu_{\rm SN}=0.06$/century. The black dot-dashed line corresponds to a naive estimate of all accelerated protons. The dotted green lines correspond to particles trapped inside the SNR and suffering adiabatic losses and released at the end of the adiabatic phase. The orange dashed line corresponds to particles constantly escaping in the ISM. The solid red line corresponds to the sum of trapped and escaping particles. The dashed blue line indicates that diffusive shock re-acceleration does not play a significant role. The data points and yellow area correspond to data from various experiments as in~\citep{cristofari2020}.}
\label{fig:LIS_oneSN}       % Give a unique label
\end{figure}

When exploring the parameter space accessible to the various SNe, one can then wonder if it may be possible to account for the entire LIS with only type of rare SNe/SNRs. We illustrate the idea of such possibility in Fig.~\ref{fig:LIS_oneSN}. One major issue with this idea is that it relies on a very unusual and rare type of SNe to account for the entire LIS - which is not a satisfying solution - that would leave almost no room for the contribution of the rest of the other SNRs, that are clearly known to accelerate high energy particles (if only because of gamma-ray observations) and are expected to contribute to the LIS. 

In short, to the question: "which sources could be pevatrons?" The answer is: 1. Probably SNRs (few and rare); 2. Many other sources could also be involved (massive stars, superbubbles, etc.). To the slightly more precise question "How to produce enough PeV protons, and a hard-enough spectrum above PeV?", we need to specify the role of various SNRs, and the parameter governing particle acceleration (efficiency, duration of the acceleration); the role played by other accelerators, and the origin of fine features (e.g., $\sim 10$ TeV bump). 

Other important issues have been also mentioned for SNRs, such as our inability to understand: a) what is the spectrum of accelerated particles? (i.e. the role of non-linear effects and deviation from the canonical test-particle spectrum $\propto p^{-4}$) b) What is the spectrum of protons/electrons \textit{released} in the ISM?~\citep{cristofari2021,cristofari_review2021,cristofari2022}.

\section{Other candidates}

Aside from the proton spectrum, the problem of $^{22}$Ne/$^{20}$Ne ratio, found to be a factor of $\approx 5$ larger in CRs than in the local ISM~\citep{binns2005}, has been interpreted as an indication that stellar winds termination shocks, and especially Wolf-Rayet winds, which are enriched in $^{22}$Ne must be involved~\citep{casse1982,prantzos2012,tatischeff2018}. 

Next to that, the detection of gamma rays from massive stars, such as the Cygnus cocoon, the central molecular zone, or Westerlund 1, with an extended profile scaling as the inverse of the radial distance from the center $\propto 1/R$~\citep{aharonian2019}, suggests a continuous injection of particles from massive stars and has helped dust off the idea that massive stars could be involved in the origin of CRs~\citep{casse1980,volk1982,cesarsky1983}.
The role of stellar clusters and interstellar bubbles, whether they are young and compact, with massive stars centered, or older with multiple supernova remnants complexifying the bubble environment has been actively studied~\citep{bykov2001,parizot2004,ferrand2010,gupta2020,vieu2022}. 

If one considers SNRs, superbubbles, and wind termination shocks (WTS) as potential Galactic CR sources, it seems that alone, none of these candidates can by itself easily solve  all the issues discussed above - energetics, injected spectrum, acceleration up and beyond PeV with the correct spectrum, 10 TeV bump, $^{22}$Ne/$^{20}$Ne ratio. The possibility for a mixed contribution from these candidates has been investigated, and some authors have for instance found, confronting available CR abundance to various benchmark scenarios,  that a mixed scenario where most of CRs come from SNRs in superbubbles and a $\sim 5$\% contribution from WTS of young stellar clusters could decently work~\citep{tatischeff2021}. 

\section{Open questions}
In addition to the major issues discussed so far, several intriguing results on CRs have yet to be explained~\citep[see e.g. detailed discussions and references in][]{gabici2019}:
\begin{enumerate}
\item the spectral hardening towards the Galactic center 
\item the small spatial gradients of CRs 
\item the very low level of anisotropy and phase pointing away from the Galactic center (<100 TeV)
\item the origin of small-scale anisotropies
\item the He spectra differing from the proton spectra
\end{enumerate}

Some of these issues might be directly connected to the problem of the interpretation of the Galactic CR spectrum around the knee and to the problem of the identification of Galactic sources. For instance, the wild hypothesis was discussed during UHECR2022,  that if the trend of He and proton spectra are extrapolated,  what is now understood to be the proton knee at $\sim 3-4$ PeV could in fact be Helium dominated. If this were in fact the case, all the above discussions should need revision, especially the discussions on the nature of the proton pevatrons.

The next years will be absolutely determinant for the search of the Galactic CR sources. Instruments capable of characterizing the gamma-ray spectra in the 100 TeV range, e.g., CTA, will be of special interest. It will not only help to clarify the role of SNRs, superbubbles, and WTS in the origin of Galactic CRs but also answer essential questions on shock acceleration (efficiency, spectra of accelerated particles, proton/electron ratio accelerated).

\begin{figure}
\centering
\sidecaption
\includegraphics[width=8cm,clip]{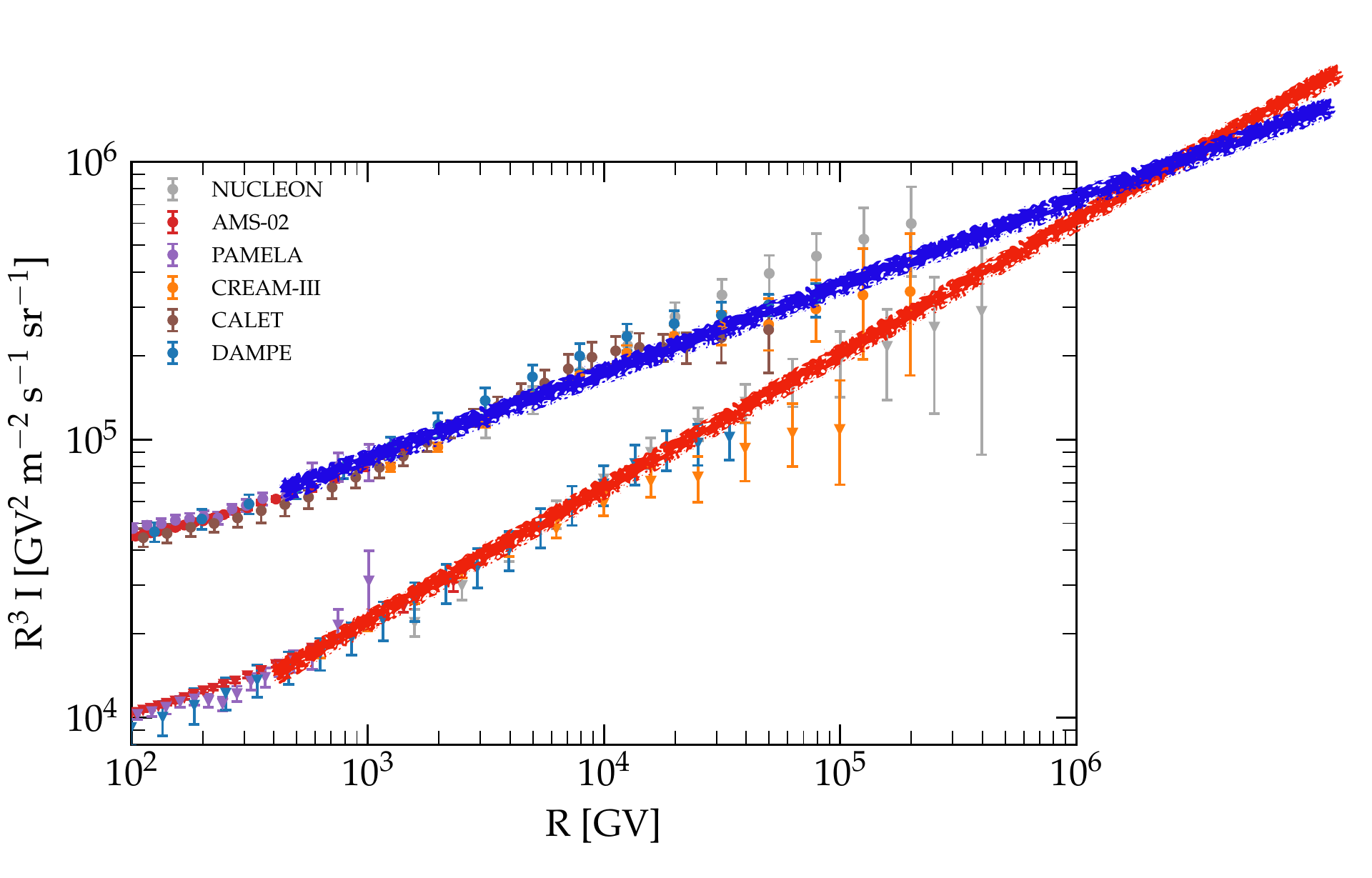}
\caption{The He and proton spectra measured by various experiments. The blue and red lines are very wild extrapolations of the proton and Helium trends to higher energies. Data compiled by C. Evoli \url{https://github.com/carmeloevoli}}
\label{fig:He_H}       % Give a unique label
\end{figure}

%Especially, since in the very--high--energy domain, the energy of hadronic gamma rays from pion production/decay is roughly one tenth of the energy of the parent protons, $E_{\gamma}= E_{\rm p}/10$, so that the detection of $\sim 100$ TeV gamma rays can help probe $\sim$ PeV protons.  Hence, a strong motivation for the hunt for pevatrons. 

% add figure Etienne? 

\section{Acknowledgement}
PC warmly thanks the organizing committee of UHECR2022, and C. Evoli for the plots.

%
% BibTeX or Biber users please use (the style is already called in the class, ensure that the "woc.bst" style is in your local directory)
% \bibliography{name or your bibliography database}

\end{document}